\documentclass[runningheads]{llncs}
\usepackage[sort,space]{cite}
\usepackage{amsmath,amssymb,amsfonts}
\usepackage{algorithmic}
\usepackage{textcomp}
\usepackage[ruled,linesnumbered]{algorithm2e}
\usepackage{multirow}
\usepackage{makecell}
\usepackage{caption}
\usepackage{booktabs}
\usepackage{outlines}
\usepackage{bbold}
\usepackage{graphicx}
\usepackage{subfigure}
\usepackage{enumitem}
\usepackage{soul}
\usepackage{color}
\usepackage{comment}
\usepackage{diagbox}
\usepackage{epstopdf}
\usepackage{wrapfig}

\def\todo#1{\textcolor{black}{#1}}

\newcommand{\eg}{{\it e.g.}}
\newcommand{\etal}{\textit{et al}.~}

\newcommand{\ie}{{\it i.e.}}

\newcommand{\sys}{\textsc{SkyMask}\xspace}

\usepackage{eccv}

\usepackage{eccvabbrv}

\usepackage{graphicx}
\usepackage{booktabs}
\usepackage{marvosym}
\usepackage[accsupp]{axessibility}  % Improves PDF readability for those with disabilities.

\usepackage{hyperref}

\usepackage{orcidlink}

\begin{document}

\title{\sys: Attack-agnostic Robust Federated Learning with Fine-grained Learnable Masks}

\titlerunning{\sys: Attack-agnostic Robust Federated Learning}

\author{Peishen Yan\inst{1}%\orcidlink{0000-0002-2433-2128} 
\and
Hao Wang\inst{2}%\orcidlink{1111-2222-3333-4444}
\and
Tao Song\inst{1}\textsuperscript{(\Letter)}%\orcidlink{2222--3333-4444-5555}
\and
Yang Hua\inst{3}%\orcidlink{2222--3333-4444-5555}
\and
Ruhui Ma\inst{1}%\orcidlink{2222--3333-4444-5555}
\and
\\Ningxin Hu\inst{4}%\orcidlink{2222--3333-4444-5555}
\and
Mohammad Reza Haghighat\inst{5}%\orcidlink{2222--3333-4444-5555}
\and
Haibing Guan\inst{1}%\orcidlink{2222--3333-4444-5555}
}

\authorrunning{P.~Yan et al.}

\institute{Shanghai Jiao Tong University, Shanghai, China \\
\email{\{peishenyan,songt333,ruhuima,hbguan\}@sjtu.edu.cn}
\and
Stevens Institute of Technology, Hoboken, USA \\
\email{hwang9@stevens.edu}
\and
Queen's University Belfast, Belfast, UK\\
\email{Y.Hua@qub.ac.uk}
\and
Intel Corporation, Shanghai, China\\
\email{ningxin.hu@intel.com}
\and
Intel Corporation, Santa Clara, USA\\
\email{mohammad.r.haghighat@intel.com}}

\maketitle

\begin{abstract} 
Federated Learning (FL) is becoming a popular paradigm for leveraging distributed data and preserving data privacy. 
However, due to the distributed characteristic, FL systems are vulnerable to Byzantine attacks that compromised clients attack the global model by uploading malicious model updates. 
With the development of \textit{layer-level} and \textit{parameter-level} fine-grained attacks, the attacks' stealthiness and effectiveness have been significantly improved.
The existing defense mechanisms solely analyze the \textit{model-level} statistics of individual model updates uploaded by clients to mitigate Byzantine attacks, which are ineffective against fine-grained attacks due to unawareness or overreaction.
To address this problem, we propose~\textsc{SkyMask}, a new attack-agnostic robust FL system that firstly leverages fine-grained learnable masks to identify malicious model updates at the parameter level. 
Specifically, the FL server freezes and multiplies the model updates uploaded by clients with the parameter-level masks, and trains the masks over a small clean dataset (\textit{i.e.}, \textit{root dataset}) to learn the subtle difference between benign and malicious model updates in a high-dimension space. 
Our extensive experiments involve different models on three public datasets under state-of-the-art (SOTA) attacks, where the results show that \sys achieves up to 14\% higher testing accuracy compared with SOTA defense strategies under the same attacks and successfully defends against attacks with malicious clients of a high fraction up to 80\%.
Code is available at \url{https://github.com/KoalaYan/SkyMask}.
\end{abstract}
\section{Introduction}
Federated learning (FL)~\cite{mcmahan2017communication} is a distributed machine learning paradigm that addresses the conflicts between ML training and data privacy. Instead of centralizing training data, FL distributes a global model to clients, trains the model locally, and aggregates the local models into a new global model without leaking any clients' local data, which has been applied in many computer vision applications, including large-scale visual classification~\cite{hsu2020federated}, object detection~\cite{liu2020fedvision}, medical imaging~\cite{guo2021multi, liu2023clip}, and many others~\cite{zhang2021federated, liu2021feddg, li2023diverse}.

However, due to FL's distributed design, attackers can easily attack the FL system by compromising client participants and smuggling malicious model updates, known as Byzantine attacks~\cite{bhagoji2019analyzing, baruch2019little, xie2019dba, fang2020local, shejwalkar2021manipulating, qureshi2021performance}. 
%
% Based on whether the compromised clients poison their local data or models, Byzantine attacks can be classified into \textit{data poisoning} and \textit{model poisoning} attacks.
%
% Data poisoning attacks corrupt the global model by injecting contaminated training samples to the local dataset~\cite{tolpegin2020data,qureshi2021performance}. 
%
% Model poisoning attacks~\cite{bhagoji2019analyzing,baruch2019little} directly manipulate local model updates to compromise the global model. 
%
To tackle Byzantine attacks, researchers have explored various defense strategies, most of which leverage coarse-grained model-level statistics %, distances, and cosine-similarities 
to detect outlier model updates~\cite{blanchard2017machine,cao2020fltrust, shejwalkar2021manipulating, rieger2022deepsight,nguyen2022flame} or greedily filter out outlier parameters~\cite{yin2018byzantine, guerraoui2018hidden}.

% \begin{wrapfigure}{r}{.355\textwidth}
%     \begin{minipage}{\linewidth}
%     \centering
%     \subfigure[The PCA of model updates]{\label{fig:toy_model}\includegraphics[width=\linewidth]%, trim=20 20 20 20,clip]
%     {Figure/FedGrad_PCA/model_pca}} \\
%     \subfigure[The PCA of masks]{\label{fig:toy_mask}\includegraphics[width=\linewidth]%, trim=20 20 20 20,clip]
%     {Figure/Skymask_PCA/mask_pca}} %.eps
%     \vspace{-0.1in}
%     \caption{Visualizing model updates and masks with PCA.}
%     \vspace{-0.15in}
%     \label{fig:toy}
% \end{minipage}
% \end{wrapfigure}

\begin{wrapfigure}{r}{.355\textwidth}

    \begin{minipage}{\linewidth}
    \centering
    \vspace{-0.25in}
    \subfigure[The testing accuracy under the fine-grained attack]{\label{fig:toy_acc}\includegraphics[width=\linewidth]%, trim=20 20 20 20,clip]
    {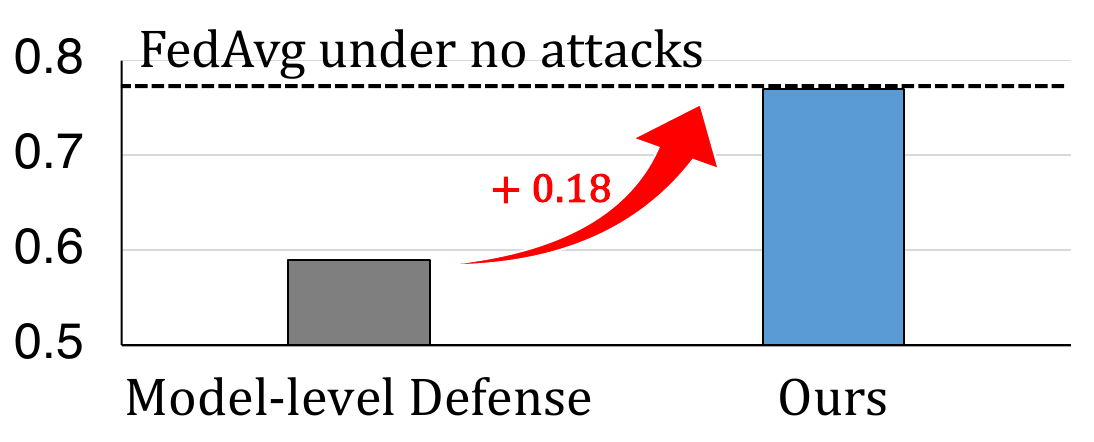}} \\
    \vspace{-0.1in}
    \subfigure[The PCA of model updates]{\label{fig:toy_model}\includegraphics[width=\linewidth, trim=0 10 0 0,clip]
    {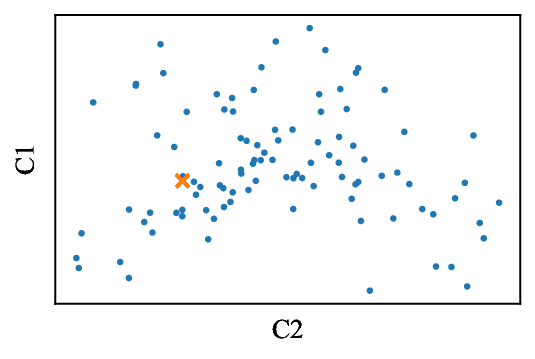}} \\
    \vspace{-0.1in}
    \subfigure[The PCA of masks]{\label{fig:toy_mask}\includegraphics[width=\linewidth, trim=0 10 0 0,clip]
    {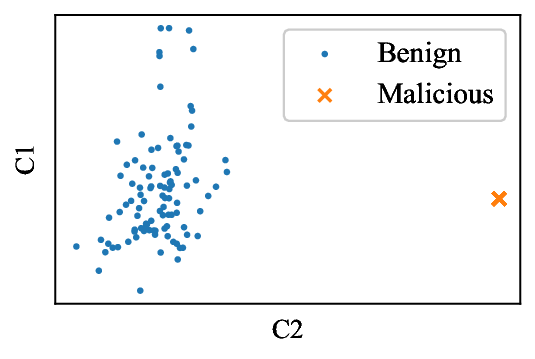}}  %.eps
    \vspace{-0.1in}
    \caption{Visualizing model updates and masks with PCA.}
    \vspace{-0.25in}
    \label{fig:toy}
\end{minipage}
\end{wrapfigure}

Unfortunately, \textit{fine-grained} attacking methodologies utilize their precise balancing between attacking stealthiness and effectiveness to circumvent existing defense mechanisms, \eg, Fang attack~\cite{fang2020local} and AGR-agnostic attack~\cite{shejwalkar2021manipulating} leverage the varying sensitivities of different neural network layers and parameters to craft adaptive and stealthy attacks. 
Existing model-level defense strategies either fail to prevent such fine-grained attacks or spill over the benign clients, which extensively sacrifices training efficiency and model quality. 
We conduct a real-world experiment to show the poor performance of the model-level defense~\cite{tolpegin2020data} under a fine-grained attack~\cite{shejwalkar2021manipulating} \todo{(detailed settings provided in supplementary \S{A.5}).} 
Existing defense method exhibits 18\% testing accuracy degradation.
We visualize model updates of all clients in a single iteration by the principal component analysis (PCA) as~\cite{tolpegin2020data}. As depicted in Fig.~\ref{fig:toy_model}, the fine-grained attack camouflages the malicious updates within benign ones, underscoring the insufficiency of analyzing local model updates solely at the model level.

% Binary masks have showcased their ability to capture models' intrinsic characteristics at the parameter-level, which are used in deep learning for model pruning~\cite{zhou2019deconstructing,li2021fedmask}. 
%
To explore whether binary masks~\cite{zhou2019deconstructing,li2021fedmask} can effectively detect malicious clients, we simultaneously train the binary masks for each model update on the server in the aforementioned toy experiment.
Fig.~\ref{fig:toy_mask} illustrates the visualization of trained masks using PCA, where masks applied to malicious model updates distinctly stand out among all masks. This highlights the capability of learnable masks to differentiate between benign and malicious model updates, and this method improves the testing accuracy by 18\%.

Based on this observation, we propose \sys, a new Byzantine-robust FL system that defends against agnostic attacks with fine-grained learnable masks at the parameter level. 
Specifically, after collecting local models from participant clients, the FL server \textit{freezes} and \textit{multiplies} each of the local model parameters with a learnable mask variable. 
Then, all masked local models are trained on a small and clean dataset (\ie, the root dataset~\cite{cao2020fltrust}) together. 
During the training process on the root dataset, the mask variables, which are applied to the frozen local model parameters, learn to improve the model accuracy. 
Particularly, the masks applied to malicious models are trained to \textit{correct} their potential misbehaviors at the parameter level. 
Since utilizing fine-grained learnable masks does \textit{not} enforce customization for specific attacks or refactoring the FL training objective functions, \sys gains three superior capabilities beyond existing defense strategies: 
1) effective and efficient detection of fine-grained stealthy attacks, 2) attack-agnostic defense against a wide spectrum of attack methods, and 3) strong compatibility and complementarity with existing FL and defense methodology. 
To sum up, our main contributions are as follows:
\begin{itemize}%[noitemsep,topsep=2pt,parsep=2pt,partopsep=0pt]
    \item We propose fine-grained learnable masks that can capture models' intrinsic characteristics at the parameter level in a uniform high-dimension space.
    \item We design \sys, a new attack-agnostic Byzantine-robust FL system, which applies fine-grained learnable masks to detect malicious clients and defend against Byzantine attacks.
    \item We empirically evaluate our \sys on various benchmarks under seven SOTA attacking methods and compare them with existing defense methods, where experimental results show that \sys achieves up to 14\% higher testing accuracy compared with existing defenses.  
\end{itemize}
\section{Preliminaries \& Related Work}

\subsection{Federated Learning}
A typical FL system includes $n$ clients and a server. Each client $i$ has a local dataset $D_i$, $i=1,\dots,n$. In each communication round $t$, the server selects a subset of the clients $\mathcal{N}_t$ to execute the following steps: 1) \textit{Model distribution}: The server distributes the global model $W_t$ to the selected clients. 2) \textit{Local training}: The clients receive global model $W_t$ as the initial model $W^i_{t+1}$ for local training, and repeat $W_{t+1}^i:=W^i_{t+1}-\beta \nabla f(W^i_{t+1};D_i)$ for $l$ local iterations, where $f(\cdot;\cdot)$ is the empirical loss function and $\beta$ denotes the local learning rate. 
The corresponding local model update is $\Delta W^i_{t+1}=W_t-W^i_{t+1}$.
%, so for the server, it is equivalent to the local model $W^i_{t+1}$. 
3) \textit{Model aggregation}: After the training, each client $i$ uploads its model weights $W_{t+1}^i$ to the server for aggregation. FedAvg~\cite{mcmahan2017communication} performs weighted model averaging to update the global model: $W_{t+1}=\sum_{i\in \mathcal{N}_t} \frac{|D_i|}{\sum_{i\in \mathcal{N}_t}|D_i|}W_{t+1}^i$, where we set $|\mathcal{N}_t|=n$ for simplicity.

\subsection{Threat Model \& Byzantine Attacks on FL}
\label{sec:background-attack}

Malicious clients compromised by the attacker participate in the FL process and have access to the knowledge of neural network models, learning rates, and objective functions. Besides, the attacker can fully control these clients' activities (\eg, modifying the model updates) and has complete access to their local datasets~\cite{fang2020local, cao2020fltrust, shejwalkar2021manipulating}. 

Existing Byzantine attacks on FL can be categorized into \textit{untargeted} and \textit{targeted} based on the adversary's goal: 
\textit{1) Untargeted attacks} seek to corrupt the global model and minimize its accuracy on any test input~\cite{tolpegin2020data,fang2020local,shejwalkar2021manipulating,baruch2019little}. Specifically, fine-grained attacks, represented by Fang attack~\cite{fang2020local} and AGR-agnostic attack~\cite{shejwalkar2021manipulating}, adaptively trade off the attacking stealthiness and effectiveness at a fine granularity as an optimization problem, which preserves the effects of poisoning with as few model-level outliers as possible.
\textit{2) Targeted attacks} aim to reduce the utility of the global model on attacker-specified tasks~\cite{bhagoji2019analyzing,sun2019can,bagdasaryan2020backdoor,xie2019dba,li20233dfed,lyu2023poisoning,nguyen2024iba,zhang2024a3fl}. % and those utilizing specific triggers are termed backdoor attacks~\cite{bagdasaryan2020backdoor,xie2019dba,li20233dfed,lyu2023poisoning,zhang2024a3fl}. 
%
% Those that reduce the utility of the global model on test inputs containing specific signals or triggers are specifically termed backdoor attacks~\cite{bagdasaryan2020backdoor,xie2019dba,li20233dfed,lyu2023poisoning,zhang2024a3fl}.

\subsection{Byzantine-robust FL Algorithms}\label{sec:background-defense}

We classify existing representative Byzantine robust algorithms into two categories based on their strategies:\\
% similarity-based 
\textbf{Model-level defense strategies.}
Some defense methods take model-level distance as a basis for determining whether a client is malicious, \eg, Euclidean distance in Krum~\cite{blanchard2017machine}, or cosine distance in DeepSight~\cite{rieger2022deepsight} and FLAME~\cite{nguyen2022flame}. {FLTrust}~\cite{cao2020fltrust} trains a root model with a small root dataset, and takes the cosine similarity between a local model and the root model as the weight in aggregation. {Tolpegin defense}~\cite{tolpegin2020data} identifies malicious model updates by employing PCA dimension reduction and clustering.
{FLDetector}~\cite{zhang2022fldetector} predicts a client's model update based on historical data and compares it with the received update. It employs consistency analysis on model updates to identify malicious clients.\\
%
\begin{comment}    
{FLTrust}~\cite{cao2020fltrust} trains a root model with a small root dataset, and takes the cosine similarity between a local model and the root model as the weight in aggregation. 
%
{Krum}~\cite{blanchard2017machine} assumes the number of malicious clients is known as $n_m$. The server computes a score for each model update according to the distance between it and the $n-n_m-2$ nearest ones. The one with the smallest score is selected as the global model update.
% 
{Tolpegin defense}~\cite{tolpegin2020data} identifies malicious clients by standardizing model updates, employing the principal component analysis (PCA) for dimension reduction, and detecting them through clustering.
%
{FLDetector}~\cite{zhang2022fldetector} employs consistency analysis on model updates to identify malicious clients. It predicts a client's model update based on historical data and compares it with the received update. Inconsistencies indicate potential malice.
%
DeepSight~\cite{rieger2022deepsight} identifies malicious clients through the analysis of their cosine distances, Normalized Update Energies and Division Differences. Subsequently, it clips the aggregated gradient to enhance defense.
%
FLAME~\cite{nguyen2022flame} detects malicious clients through cosine distance analysis and subsequently utilizes a customized weak DP approach, integrating noise boundary proofing and dynamic clipping bound.
\end{comment}
%
\textbf{Greedy parameter-level filtering strategies.}
{Trimmed-Mean (Trim)}~\cite{yin2018byzantine} is a coordinate-wise aggregation rule assuming the number of malicious clients $n_m$ is known, removes the parameters of the smallest and largest $n_m$ values, and averages the remaining.% Median retains the corresponding coordinate's median value as the new global model's value. 

However, existing robust FL strategies can hardly defend against those fine-grained attacks without sacrificing training efficiency and model quality by utilizing coarse-grained detection or greedy parameter-wise filtering, which motivates us to explore malicious client detection with fine-grained learnable masks.

\subsection{Root Dataset}
Considering that the server has no root of trust to decide whether a model update is malicious or not, FLTrust~\cite{cao2020fltrust} is the first to introduce the concept of root dataset.
It can take advantage of the root dataset even if the distribution diverges from the overall data distribution or the size is less than one hundred, which is easy for the server and FL manager to manually collect or generate a small root dataset. 
% %
The rationale for utilizing root datasets has been validated in numerous other studies~\cite{guo2021siren,miao2022privacy,wang2022protect,li2023martfl}. \todo{We evaluate the impact of the root dataset's data distribution in supplementary \S{C.1}}.
\section{Methodology}

% \subsection{Goals} 

% We aim to design a server-based robust FL system. The server can collect or generate a small representative root dataset. 
We aim to design a server-based robust FL system. The server has a small representative root dataset following previous work~\cite{cao2020fltrust}. 
Moreover, this root dataset can be imbalanced. % \todo{(analyzed in supplementary \S{C.1}).} 
%
% We allow reasonable additional computation costs and extra storage space to defend against Byzantine attacks. We aim to detect all the malicious model updates and filter them out at each communication round to eliminate the impact of attacks. 
% 
% We should also keep as many benign clients as possible in the filtered FL client set to train a final global model with sufficient local data for satisfactory accuracy. 
%
For the proposed Byzantine-robust FL system, it should promise the following three features:
(1) \textbf{Robustness}. The system should preserve the global model accuracy under attacks. In particular, we should maintain a high malicious client detection accuracy and a low misidentification rate of benign ones to eliminate the impact of attacks.
(2) \textbf{Generality}. The system should also stay effective on different datasets and model structures under different attacks.
(3) \textbf{Efficiency}. The defense system is designed on the server instead of resource-constrained client. Besides, when performing malicious client detection and defense on the server, we allow reasonable additional computation costs and extra storage space.
%
% (3) \textbf{Scalability}. The system should support a large scale of clients and be able to cope with a high fraction of malicious clients.

% \begin{itemize}[noitemsep,topsep=2pt,parsep=2pt,partopsep=0pt]
%     \item \textbf{Robustness}. The system should preserve the global model accuracy when attacked by malicious clients. In particular, we should maintain a high malicious client detection accuracy and a low misidentification rate of benign ones (\ie, false positive rate). The system should also stay effective on different datasets and model structures under different attacks.
%     \item \textbf{Efficiency}. The system should not introduce extra computation and communications overhead to the clients, which are often resource-constrained devices in FL. Besides, when performing malicious client detection and defense on the server, the algorithm should only incur a proper complexity. 
%     \item \textbf{Scalability}. The system should still support a large scale of clients and work under a high fraction of malicious clients (\eg, 80\%) since FL systems usually involve a large number of client devices. 
% \end{itemize}

\subsection{Overview}
\label{sec:ov}

\begin{figure}[tb]
    \centering
    % \vspace{-0.15in}
    \includegraphics[width=\linewidth]{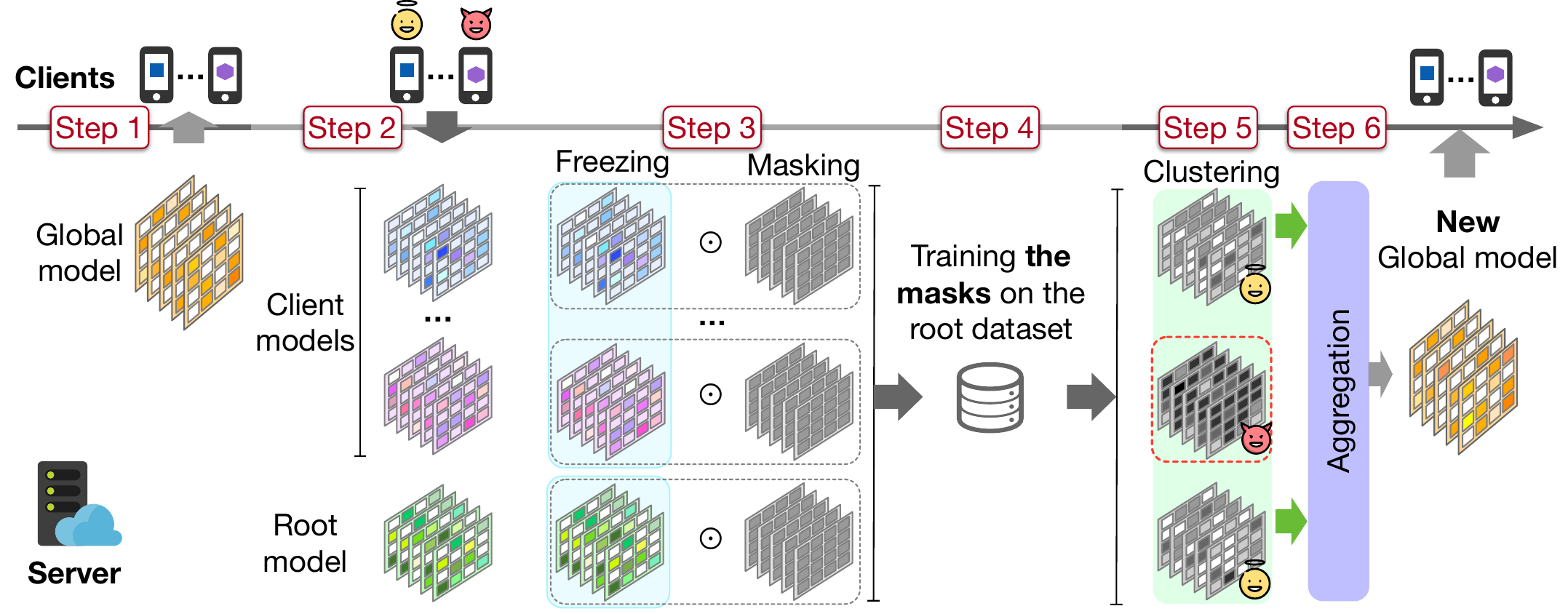}
    % \vspace{-0.05in}
    \caption{\sys's workflow.}
    \vspace{-0.15in}
    \label{fig:framework}
\end{figure}

When initializing the FL process, the server creates an initial global model and builds a root dataset as~\cite{cao2020fltrust}. Fig.~\ref{fig:framework} presents \sys's workflow, where each communication round has six steps: 
\textbf{Step 1}, the server distributes the global model parameters to each client; 
\textbf{Step 2}, the clients load the parameters, train the local models, and send back these model updates; % The server also conducts local training on the root dataset to generate a root model (root model).
\textbf{Step 3}, the server freezes and multiplies each local model parameter with a learnable mask variable; 
\textbf{Step 4}, all the masked local models are trained together on the root dataset to converge; 
\textbf{Step 5}, the server detects and removes the malicious clients by clustering all trained masks; 
\textbf{Step 6}, the server aggregates the remaining model updates into a new global model. 

The key component of \sys is the learnable mask-driven detection. Specifically, \sys constructs a same-size fine-grained mask for each local model update and trains these masks by optimizing the aggregation result of masked local models until they converge. 
Byzantine attacks take impairing global model performance as an optimization objective, so the poisoned model updates tend to make the global model far from normal behavior. 
Therefore, to correct potential misbehaviors of malicious models, the masks applied to the malicious models learn to mitigate the side effects of poisoned parameters through training on the root dataset.
In this way, the learnable masks form a high-dimensional representation space to capture the characteristics of both malicious and benign model updates. Then, the clusters formed by masks are the basis for determining whether a client is malicious. After removing all the detected malicious model updates, \sys calculates the average of remaining model updates and uses it to update the global model.

\subsection{\textsc{SkyMask} Algorithm}
\label{sec:detect_strat}
Most existing defense strategies are based on the local model updates' coarse-grained statistics or greedily filtering out outlier parameters. However, fine-grained attacks corrupt a small set of particular layers or parameters to work around existing defense strategies easily. 
Thus, we apply fine-grained learnable masks and clustering analysis to detect and defend against malicious clients. Our defense strategy has two stages: 1) mask initialization and training stage; 2) mask clustering and classification stage.

\subsubsection{\textbf{Mask initialization and training.}} 

In the first stage, the server freezes all the model updates and assigns a learnable mask $m_i$ of the same size as the model for each client $i$ and initializes them all with $1$s. The model aggregated from masked local models $\tilde{W}_{t+1}$ (called aggregated masked model) is computed by averaging the masked local models $m_i \odot W^i_{t+1}$. Then, the masks start training on the root dataset with frozen local models. % as Fig.~\ref{fig:framework} Step 4. 
At the beginning of iteration $t+1$, $\tilde{W}_{t+1}$ is calculated as:
% 
% \begin{equation*}
% \end{equation*}
% \begin{equation}
% \tilde{W}_{t+1}=\sum^n_{i=1}\frac{1}{n} m_i \odot W^i_{t+1}, \text{where }W^i_{t+1}=W_t-\Delta W^i_{t+1}. \label{eq:training_model}
% \end{equation}
\begin{equation}
\tilde{W}_{t+1}=\sum^n_{i=1}\frac{1}{n} m_i \odot W^i_{t+1}. \label{eq:training_model}
\end{equation}

We train the aggregated masked model $\tilde{W}_{t+1}$ on the root dataset. Then, the update process back-propagates gradients with the help of $\tilde{W}_{t+1}$ to masks and applies a gradient descent algorithm to the masks. 

If there is no constraint on the value range of mask variables during the mask training, some masks can have very large values, while others can be so small for the same dimension when these masks converge. %The masks are ranged $(-\infty, +\infty)$. %It is hard to detect malicious clients by the masks ranged $(-\infty, +\infty)$ since the difference in magnitude of masks prevents effectively characterizing the local model. 
Huge differences in magnitude between masks  can affect the detection of malicious clients. 
Therefore, we redesign the masks to show the contribution of each local model update in the optimized model.

We use binary masks to extract the parameter-level characteristics of local models because binary can directly determine whether a parameter of model update participates in the global model.
Since binary parameters are not derivable, we approximate the binary masks in training by setting the mask with real values and applying a sigmoid function $\sigma(\cdot)$ instead of a hard threshold to reduce the enormous gradient variance~\cite{li2021fedmask}. Furthermore, it is easy for the gradient to be back-propagated to the real-value masks. The approximation of a binary mask is calculated as $\tilde{m}_i=\sigma(m_i)$, which is ranged $(0,1)$. %After the real-value masks converge, the final results of binary mask $\hat{m}_i$ are computed by thresholding according to a threshold $\tau$ and the trained mask $\tilde{m}_i$. Denoting the parameter located in $k$th dimension of binary mask as $\hat{m}_i[k]$, if $\tilde{m}_i[k] > \tau$, we set $\hat{m_i} [k]=1$; otherwise, we set $\hat{m_i} [k]=0$.

Since the mask's limitation is changed, the original Equation~\ref{eq:training_model} is unsuitable. To adapt for the approximated binary masks, $\tilde{W}_{t+1}$ is re-written as:
\begin{equation}
\tilde{W}_{t+1}=\frac{\sum^n_{i=1}\tilde{m}_i \odot W^i_{t+1} }{ \sum^n_{i=1}\tilde{m}_i}.
\label{eq:til_mask}
\end{equation}

% The optimization process prefers to reject the parameters of the poisoned local model and sets the corresponding parameters of the mask as $0$.

It can be formulated as an element-wise weighted averaging algorithm. If the parameters of a poisoned local model are toxic to the aggregated masked model $\tilde{W}_{t+1}$, the mask tends to reduce the involvement of these parameters to $\tilde{W}_{t+1}$ in the optimization phase.  Thus, the 0-1 pattern in the mask can represent whether the corresponding local model is malicious. 

The mask training task is not different from the main task in FL except for the variables. The objective function is:
\begin{equation}
    f(\tilde{W}_{t+1},D_{r})=\sum\limits_{(x,y)\in D_r} L(\text{output}(x,\tilde{W}_{t+1}),y).
\end{equation}
The masks are updated in each mask training iteration as follows:
\begin{equation}
    m_i := m_i -\gamma\cdot \nabla_{m_i} f(\tilde{W}_{t+1},D_{r}),
\end{equation}
where $\gamma$ represents the mask learning rate.
This process will not stop until these masks converge. After the real-value masks converge, the final results of binary mask $\hat{m}_i$ are computed as:
\begin{equation} \label{eq:thres}
    \hat{m_i} [k] = \begin{cases} 1, & \tilde{m}_i[k] > \tau \\
    0, & \tilde{m}_i[k] \leq \tau 
    \end{cases},
\end{equation}
where $\hat{m}_i[k]$ represents the parameter located in $k$th dimension of binary mask $\hat{m}_i$ of client $i$ and $\tau$ is a threshold.

\subsubsection{\textbf{Mask clustering and classification.}} 
These binary masks can sense potential attacks at a fine granularity and represent the parameter-level characteristics of each client's model update. 
% We use the PCA algorithm to reduce the dimensionality to $d$. Since the maximum dimensionality that the masks can be reduced to is limited by the number of masks $n$ and the dimensionality of masks $V$, we set $d=2$, which is the minimum possible number of clients in the FL system.
Then, we apply the Gaussian mixture model for clustering and classification. If there are no attacks, the clustering result is only one cluster, and the server aggregates all the benign local models.

To determine which cluster is benign, \sys introduces a trusted root model $W^r_{t+1}$, initialized by $W_t$ and ``locally'' trained $l$ iterations on the root dataset as other clients:
\begin{equation}
    W^r_{t+1} := W^r_{t+1} - \beta \nabla f(W_t,D_r).
\end{equation}

The server lets it join in the model set before the mask training stage, assigns one mask for it and trains this mask together with other masks.
When all the masks converge, the clients represented by the masks in the same cluster as the mask corresponding to the trusted root model are considered benign. In this way, \sys can handle various malicious fractions from zero to very high. In order not to interfere with the global model convergence, we should emphasize that the trusted root model is only used for malicious model detection and is not included in the benign model aggregation phase.

%If the fraction of malicious clients is lower than 50\%, the system can treat the larger cluster in two clusters as the benign client set.

\sys plays $R$ rounds for server-clients communication, and each round contains six steps, as \S\ref{sec:ov} describes. The critical point of \sys is our fine-grained malicious client detection strategy mentioned above, which can provide the remaining benign client set $U_b$ for aggregating a new global model. Given the aggregation algorithm specified as FedAvg, the global model for the next communication round $W_{t+1}$ is calculated as:
\begin{equation}
    W_{t+1} = W_t - \alpha\cdot \sum\limits_{i\in U_b} \frac{|D_i|}{\sum_{i\in U_b} |D_i|}\Delta W^i_{t+1}, 
\end{equation}
where $\alpha$ denotes the global learning rate.
%
% It can even enhance the existing robust aggregation algorithms.
%
If it completes the malicious detection task, removing only the malicious model updates and leaving the benign model updates untamed, its convergence will be consistent with the original aggregation algorithm.
%
% It should be noted that \sys is more concerned with malicious client detection and can be added as an additional module to the aggregation algorithm.
\sys serves as an efficient method for identifying malicious clients within an FL framework. Its modular design allows for integration as an adjunct component into diverse aggregation algorithms, ensuring compatibility across a spectrum of existing aggregation techniques.

% \begin{figure}[t]
%     \centering
%     \includegraphics[width=.95\linewidth]{Figure/mask_Work/mask_work.eps}
%     % \vspace{-0.05in}
%     \caption{Visualize the training process of masks applied to malicious and benign models.}
%     % \vspace{-0.15in}
%     \label{fig:mask_work}
% \end{figure}

\subsection{Why \sys Works}
\begin{wrapfigure}{r}{.5\textwidth}
\centering
    \vspace{-0.7in}
    \includegraphics[width=0.95\linewidth]{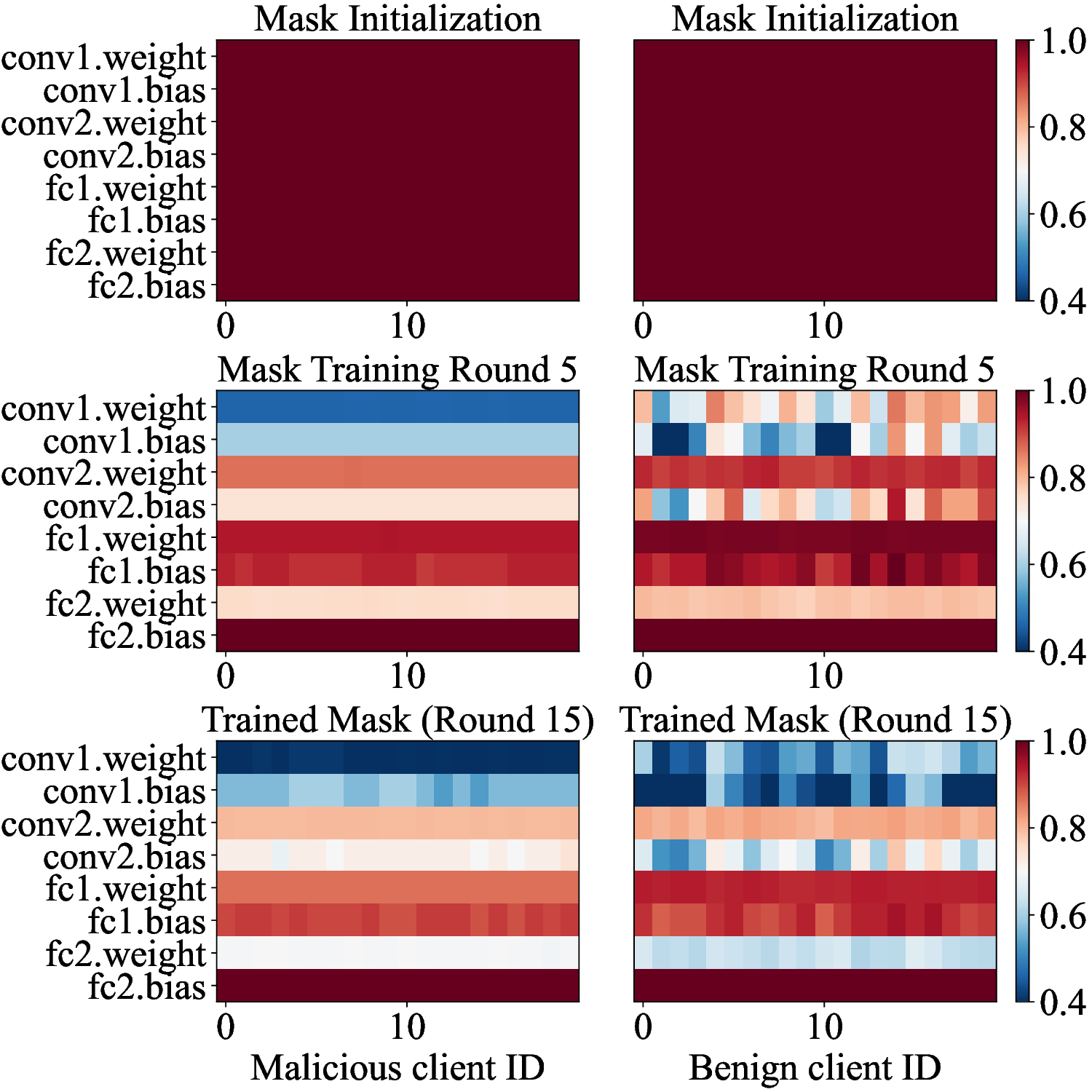}
    \vspace{-0.05in}
    \caption{The visualization of the training process of masks applied to malicious and benign models.}
    \vspace{-0.2in}
    \label{fig:mask_work}
\end{wrapfigure}
% We conduct an example experiment to show how \sys works. 
We choose a four-layer CNN model and Fashion-MNIST dataset, and set 20 of 100 clients as malicious. Under Fang-Trim attack, we take the mask training process in one communication round. We calculate the proportion of ``1''s in each part of the binary mask. A smaller value means a fewer parameters of this client are selected for the aggregated masked model. For simplicity, here only shows 20 masks for malicious and 20 for benign clients in each mask training iteration.

As shown in Fig.~\ref{fig:mask_work}, during the mask training process, the binary parameters in the ``conv1.weight'', ``conv2.weight'', and ``fc1.weight'' parts of malicious model updates are set to zeros, which are more often than those of benign updates. 
It is related to the fact that the attack prefers to poison the parameters in specific parts.
%the binary parameters in the ``conv1.weight'', ``conv2.weight'', and ``fc1.weight'' parts of malicious model updates are set to zero more often than those of benign updates. 

%
The masks gradually reduce the poisoning effect of the toxic parameters during the training so that we can classify malicious clients with the help of the masks. Therefore, \sys can perceive the poisoning attacks at the parameter level in the model updates without knowing the details of the attack, which makes it attack-agnostic.

\subsection{\sys's Complexity}
Let $O(T)$ denote the computational complexity of one local training iteration, and the model parameters' dimensionality is $V$.

The computational complexities of existing defense methods are as follows: $O(nV)$ for FLTrust, $O(nV\log n)$ for Trim, $O(n^2V)$ for Krum, $O(T+nV)$ for DeepSight, and $O(nV)$ for FLAME.
For \sys, to update the masks once, the server computes the gradient of each mask from the gradient of the aggregated masked model $\tilde{W}_{t+1}$ by backpropagation, with a computational complexity of $O(nV)$.
Hence, the total computational complexity of \sys is
$O((T+nV) t_m)+O(nV)=O(Tt_m+nV t_m)$, where $t_m$ is the number of mask training convergence iterations. Since there is no data dependency between any mask parameters, we can use parallel training~\cite{dean2012distribute} to accelerate the mask optimization process so that the time overhead can be further reduced, trading computation power for time.
The above analysis shows that the additional cost of the mask learning mechanism over other defense methods is equivalent to the cost of conducting several iterations of local training, which is acceptable compared to the computational power of the server.

% The space complexity of existing FL defense algorithms is either $O(n+nV)$ or $O(n^2+nV)$.
The space complexity of existing defense methods are: $O(nV)$ for Trim, $O(n+nV)$ for FLTrust, $O(n^2+nV)$ for Krum, DeepSight, and FLAME. For \sys, the additional space complexity is proportional to the number of clients $n$ and the space complexity of the global model parameters because \sys assigns a mask for each client and the masks have the same dimensionality as the global model. Considering that the server stores $n$ local models' parameters in the current round, the total space cost is about twice the others, and its space complexity is $O(nV)$, which is reasonable.

\section{Experiments}
% We implement a prototype of \sys based on PyTorch and 
We evaluate our method with existing Byzantine-robust FL aggregation algorithms and two malicious client detection algorithms. To verify the generalization of our method, we conduct experiments using various popular datasets paired with different models. We evaluate our method with seven popular attacks, explore the impact of the fraction of malicious clients, and verify its scalability. 
In sensitivity analysis, we discuss the influence of binarization threshold, the number of features after dimensionality reduction, and root dataset's data distribution on our method's performance \todo{(details in supplementary \S{C})}, respectively.
%
% We repeat each experiment three times and average the results. 
% All experiments run on a cloud server with eight NVIDIA RTX 2080 GPUs.

% On Fashion-MNIST and CIFAR-10 datasets, \sys improves the testing accuracy by 1--9\% compared to the best results achieved by other algorithms under the same attacks. More significantly, our method gains a 14\% improvement in testing accuracy on CIFAR-100 dataset under Min-Max attack.

\subsection{Experiment Settings}

\subsubsection{Datasets and models.}
\label{sec:dis-dataset}
We evaluate \sys on three widely-used datasets: Fashion-MNIST, CIFAR-10, and CIFAR-100~\cite{krizhevsky2009learning}. % for evaluation, which have been widely used in prior FL studies. 
Following previous methods~\cite{fang2020local,cao2020fltrust}, we construct non-IID local dataset with bias probability $p=0.5$ for each client and the root dataset for the server. We construct a CNN with two convolutional layers and two linear layers as the global model for Fashion-MNIST dataset. For CIFAR-10 and CIFAR-100 datasets, to verify \sys's effectiveness on large models in FL, we use a more complex model, ResNet20. 

\subsubsection{Baseline attacks and defenses.}
The experiments cover \textit{1) untargeted attacks}: Label-flipped (LF)~\cite{tolpegin2020data} and four fine-grained attacks, including Fang attacks~\cite{fang2020local} (Fang-Trim and Fang-Krum attacks) and two types of the AGR-agnostic attack~\cite{shejwalkar2021manipulating} (Min-Max and Min-Sum attacks);
\textit{2) targeted attack}: Scaling attack~\cite{bagdasaryan2020backdoor} and DBA~\cite{xie2019dba}.
%
% \subsubsection{Baseline defenses}
In \S\ref{sec:mask_acc}, to show the defense effectiveness, we compare \sys with five Byzantine-robust FL aggregation algorithms: FLTrust~\cite{cao2020fltrust}, Trim~\cite{yin2018byzantine}, Krum~\cite{blanchard2017machine}, DeepSight~\cite{rieger2022deepsight} and FLAME~\cite{nguyen2022flame}. In \S\ref{sec:mask_sig}, to reveal the significance of learnable masks in malicious client detection, we compare \sys with two malicious client detection algorithms: Tolpegin defense~\cite{tolpegin2020data} and FLDetector~\cite{zhang2022fldetector}.
%
% (details in supplementary material.)

\subsubsection{FL parameter settings.} Following the previous study~\cite{cao2020fltrust}, we use 100 clients in all experiments and apply multiple-client attacks. The default fraction of malicious clients is set to 20\%. % The default binarization threshold $\tau$ is $0.5$. %, and the root dataset bias probability $p=0.1$.
%
% We set both global model learning rate $\alpha$ and local model learning rate $\beta$ to $0.5$ in all experiments. For mask learning rate $\gamma$, we set $0.5$ in all three datasets. In CIFAR-10 and CIFAR-100, we allow the system to execute server-client communication for $500$ rounds and let clients do $l=5$ local training iterations per round. While in Fashion-MNIST, the system runs $2,500$ communication rounds, and each client only does a single local training iteration per round. 
%
In our experiments, the root dataset has 100 samples and no intersection with the training and test datasets, which is similar to FLTrust~\cite{cao2020fltrust}. %100 is enough for our defense, and it is easy for the server to collect. So we do not discuss the influence of the size of the root dataset. 
\todo{Other hyperparameters are detailed in supplementary \S{A.4}.}

\subsubsection{\textsc{SkyMask} variants.} \label{sec:schemes}
Intuitively, if the fraction of malicious clients is less than 50\% in the real world, we can choose the larger one in two clusters of masks as benign. This way, we can further reduce the computational pressure on the server. To verify the effectiveness with or without the trusted root model, we consider the following two schemes: 1) \textbf{\textsc{SkyMask}-NR} (no trusted root model). The FL system does not generate a trusted root model, and after all the masks are converged, the system chooses the larger one in two clusters of the masks as benign for aggregation; 
2) \textbf{\textsc{SkyMask}}. The FL system uses the defense strategy fully following the default steps described in \S\ref{sec:detect_strat}.

\subsubsection{Metrics.}\label{sec:metrics} The \textbf{testing accuracy} is obtained from the prediction accuracy of the global model on the test dataset. The \textbf{attack success rate} under targeted attack is the fraction of the trigger-embedded samples in the test dataset that are identified as the target label.
To show the malicious client detection ability, we record \textbf{false positive rate (FPR)} and \textbf{false negative rate (FNR)} in the malicious client detection stage. FPR indicates the number of benign clients that are misidentified as malicious clients divided by the total number of benign clients, and FNR indicates the number of malicious clients that are misidentified as benign clients divided by the total number of malicious clients. 
The Mean FPR/FNR is the average of FPR/FNRs in each communication round.

\subsection{The Defense Effectiveness of \textsc{SkyMask}}
\label{sec:mask_acc}
We first certificate the defensive capability under a low fraction of attacks. Then, there is also possibly no attacker in the FL system, so the defense methods should not affect the global model's performance. Eventually, we consider an extreme condition that the fraction of malicious clients is high. 

\begin{table}[t]
% \vspace{-0.1in}
\caption{FL testing accuracy under different attacks and attack success rates of targeted attack. The experimental results of targeted attacks are in the form of ``testing accuracy/attack success rate.''}
\vspace{-0.05in}
\label{tab:main}
    \centering
    \scriptsize
    \scalebox{0.75}{
    \begin{tabular}{|c|c|cccccccc|}
    \hline
    \makecell{Dataset\\(Model)}  &  \makecell{Attack} & FedAvg & FLTrust & Trim & Krum & DeepSight & FLAME & \textbf{\sys-NR} & \textbf{\sys} \\ \hline

\multirow{8}{*}{\makecell{Fashion\\-MNIST\\(CNN)}} 
& None  &\textbf{0.89}& \textbf{0.89}& 0.88& 0.83& 0.88& 0.87& \textbf{0.89}& \textbf{0.89}\\
& LF&    0.84& 0.86& 0.84& 0.83& \textbf{0.89}& 0.87& \textbf{0.89}& \textbf{0.89}\\
& Min-Max& 0.58& \textbf{0.89}& 0.70& 0.83& 0.64& 0.65& \textbf{0.89}& \textbf{0.89}\\
& Min-Sum& 0.80& \textbf{0.89}& 0.73& 0.47& 0.82& 0.75& \textbf{0.89}& \textbf{0.89}\\
& Fang-Trim& 0.42& \textbf{0.89}& 0.67& 0.82& 0.75& 0.85& \textbf{0.89}& \textbf{0.89}\\
& Fang-Krum& 0.86& \textbf{0.89}& 0.84& 0.47& 0.70& 0.78& \textbf{0.89}& \textbf{0.89}\\
& Scaling & 0.80/0.21 &	\textbf{0.89/0.08} &	0.87/0.13 &	0.82/0.09&	0.88/0.10 &	0.87/0.10  &	0.89/0.10  &	0.89/0.10 \\
& DBA & 0.86/0.51	& 0.88/0.18	& 0.88/0.31	& 0.83/0.11	& 0.89/0.16	& 0.89/0.11	& 0.89/0.11	& \textbf{0.89/0.10}\\

\hline \hline

\multirow{8}{*}{\makecell{CIFAR-10\\(ResNet20)}}
& None& \textbf{0.77}& 0.75& \textbf{0.77}& 0.54& 0.72& 0.72& 0.76& 0.76\\
& LF&   0.71& 0.74& 0.71& 0.55& 0.69& 0.68& 0.75& \textbf{0.76} \\
& Min-Max& 0.58& 0.68& 0.70& 0.52& 0.63& 0.55& \textbf{0.77}& \textbf{0.77}\\
& Min-Sum& 0.66& 0.72& 0.74& 0.32& 0.66& 0.68& 0.75& \textbf{0.77}\\
& Fang-Trim& 0.10& 0.68& 0.19& 0.52& 0.53& 0.49& 0.75& \textbf{0.76}\\
& Fang-Krum& 0.58& 0.75& 0.48& 0.19& 0.36& 0.40& \textbf{0.77}& \textbf{0.77}\\
& Scaling & 0.10/1.00 & 0.74/0.13 & 0.72/0.51 & 0.53/0.08 & 0.73/0.10 & 0.73/0.10 & \textbf{0.77/0.09} & 0.77/0.10\\
&DBA & 0.72/0.99	& 0.76/0.88	& 0.76/0.97	& 0.56/0.09	& 0.77/0.16	& 0.77/0.14 & 0.77/0.11 & \textbf{0.77/0.10}\\

\hline \hline

\multirow{8}{*}{\makecell{CIFAR-100\\(ResNet20)}}
& None	& \textbf{0.44} 	& 0.39 	& 0.43 	& 0.17 	& \textbf{0.44} 	& \textbf{0.44} 	& \textbf{0.44} 	& \textbf{0.44} \\
& LF	& 0.41 	& 0.39 	& 0.38 	& 0.11 	& 0.43 	& 0.43 	& \textbf{0.44} 	& \textbf{0.44} \\
& Min-Max	& 0.16 	& 0.30 	& 0.16 	& 0.05 	& 0.16 	& 0.19 	& \textbf{0.44} 	& \textbf{0.44} \\
& Min-Sum	& 0.33 	& 0.28 	& 0.33 	& 0.17 	& 0.35 	& 0.34 	& \textbf{0.44} 	& \textbf{0.44} \\
& Fang-Trim	& 0.01 	& 0.34 	& 0.04 	& 0.15 	& 0.20 	& 0.22 	& \textbf{0.44} 	& \textbf{0.44} \\
& Fang-Krum	& 0.03 	& 0.37 	& 0.04 	& 0.03 	& 0.03 	& 0.02 	& \textbf{0.44} 	& \textbf{0.44} \\
& Scaling	& 0.01/1.00	& 0.36/0.30	& 0.43/0.98	& 0.14/0.00	& 0.44/0.04	& 0.44/0.08	& \textbf{0.44/0.01}	& \textbf{0.44/0.01 }\\
& DBA & 0.37/0.98 &	0.40/0.90 &	0.44/0.94 &	0.16/0.01 &	0.45/0.16 &	0.45/0.15 &	 0.44/0.02 &	\textbf{0.44/0.01}\\
\hline

\end{tabular}
}
\vspace{-0.15in}
\end{table}

\subsubsection{Under a low fraction of attacks.} By default, we set the fraction of malicious clients to 20\% following the default setting in \cite{cao2020fltrust}. Table \ref{tab:main} shows that the testing accuracy of most existing defenses is reduced by more than 10\% with at least one attack. FLTrust is a relatively strong baseline defense because it achieves similar good performance as ours in a few Fashion-MNIST experiments. While under LF attack, FLTrust has a 3\% accuracy loss. Our \sys achieves the highest testing accuracy on Fashion-MNIST under different attacks compared to all others. % Scaling attack does not show an effective impact on all the defense methods. 

To demonstrate the generalization, we experiment on CIFAR-10 and CIFAR-100 dataset using a larger model, ResNet20. In such scenarios, multi-client fine-grained attacks are much more powerful. Trim, Krum, DeepSight, and FLAME are out of defense. FLTrust does not perform as well as it does on Fashion-MNIST dataset, with a severe drop in accuracy under various attacks. In contrast, our \sys reaches the highest accuracy under all attacks. Especially on CIFAR-100 dataset, our method achieves the most significant gap with other algorithms in Min-Max attack experiments. The accuracy is up to $0.44$, while the accuracy results obtained by others are less than $0.3$. Besides, \sys achieves the best and unattacked-level performance under the targeted attacks.

% Since the samples with target label 0 are 10\% of the test dataset in FashionMNIST and CIFAR-10, the attack success rate is about 10\% if there is no attack. We can see that FedAvg is overwhelmed and Trim suffers from the Scaling attack severely on CIFAR-10. \sys still keeps the robustness and performs well.

Due to the attackers' optimization for the abnormality degree, only some particular parameters of the poisoning models changed with emphasis. The results show that the four fine-grained attacks (Min-max, Min-Sum, Fang-Trim, and Fang-Krum attacks) bypass the existing defense methods in many cases.

In all these experiments, \sys not only achieves the highest testing accuracy but reaches the same accuracy level as unattacked FedAvg's, \ie, the same accuracy or no more than 1\% accuracy loss.

As \S\ref{sec:schemes} describes, the server does not need to consume resources to generate a root model if the fraction of malicious clients is less than 50\%. In the 20\% case, as shown in Table \ref{tab:main}, \sys-NR achieves good results under any attack and on any dataset, which shows the same defensive capability as \sys. Therefore, if the server confirms that the number of malicious clients is less than half of the total number, it can choose \sys-NR as the defense strategy to reduce the overhead while maintaining the same defensive capability. For a more general scenario, it can choose \sys as the defense strategy.

\subsubsection{Under no attack.} Some attacks send poisoned models to the server only at certain rounds, and some model updates can be mistaken for malicious by existing defense methods due to data heterogeneity. The system collapses if the global model performance suffers from the defense, even when there is no attack. So when there is no attack, the expected result is that any method should not affect the performance achieved by the basic aggregation algorithm FedAvg.

In Table \ref{tab:main}, almost all the existing robust aggregation algorithms affect the testing accuracy in the no-attack case. For instance, FLTrust has an accuracy loss of about 2\% on CIFAR-10 dataset and 5\% on CIFAR-100. Krum algorithm has an accuracy loss ranging from about 6\% to 27\% on different datasets.
In contrast, our \sys maintains a comparable performance as unattacked FedAvg on all datasets with a difference of less than 1\%. The existing robust aggregation algorithms either choose a subset of model updates that seem benign or use model updates' statistics to correct the impact. Hence, it is possible to introduce harmful impacts without attack. If there is no attack, our masks form only one cluster, or there are only several outliers. So \sys selects most of clients and has the same or very similar convergence compared with FedAvg. % in both IID and non-IID cases.

\begin{figure}[t]
    \centering
    \includegraphics[width=0.95\textwidth, trim= 10 10 0 0, clip]{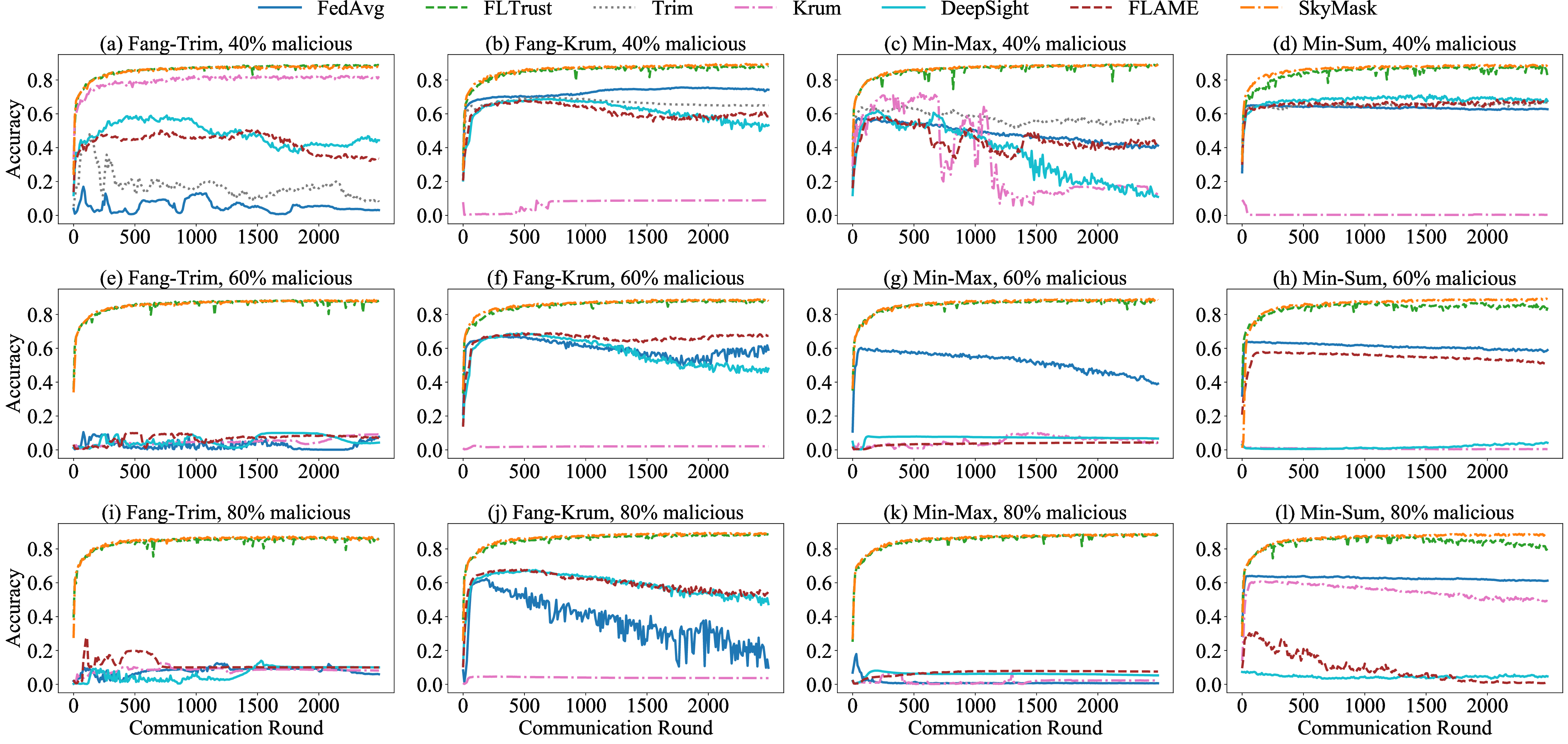} %
    \vspace{-0.1in}
    \caption{The impact of high fractions of malicious clients under fine-grained attacks.}
    \label{fig:fraction}
    \vspace{-0.1in}
\end{figure}

\subsubsection{Under a high fraction of attacks.} The experiments mentioned in the previous sections set the fraction of malicious clients to 20\%, while in a real application environment, an attacker can control a larger fraction of malicious clients for the attack. Therefore, to verify the defensive capability of our system in this case, we use four-layer CNN as the global model on Fashion-MNIST dataset and conduct comparison experiments with 40\%, 60\%, and 80\% malicious fractions. 
% Since Trim requires that the proportion of malicious clients should be less than 50\%, so it is only included in 40\% experiments. 
We present the results pertaining to fine-grained attacks in Fig.~\ref{fig:fraction}. \todo{Additional results can be found in the supplementary \S{B.1}.} 

%The experimental results are shown in Figure~\ref{fig:fraction}. 
Only \sys and FLTrust maintain the defensive capability under fine-grained attacks.
The other defense algorithms completely lose their defensive capability, exhibiting a significant gap from the unattacked level.
%
%When the fraction of malicious clients is 40\%, Trim Krum can reduce the impact of attacks under Fang-Trim attacks, but the accuracy loss is still large. Although the accuracy of FLTrust fluctuates in some communication rounds, it still tends to converge to an unattacked level. 
From Fig.~\ref{fig:fraction}(b), Fig.~\ref{fig:fraction}(c), and Fig.~\ref{fig:fraction}(d), we can see that \sys converges more stable than FLTrust when they are attacked by Fang-Krum, Min-Max, and Min-Sum attacks. In Fig.~\ref{fig:fraction}(d), the accuracy obtained by \sys is 2\% higher than FLTrust's under Min-Sum attacks.

When the fraction is more than 50\%, only \sys maintains the defensive capability. Even as the best-performing baseline defense, FLTrust's performance worsens as the fraction rises. FLTrust's testing accuracy fluctuates more, and the global model even converges in the wrong direction, \eg in Fig.~\ref{fig:fraction}(l), the accuracy obtained by FLTrust decreases from 85\% to 81\% after a sharp fluctuation. Under other attacks, the increasing fraction influences FLTrust's convergence speed, whereas \sys remains unaffected and converges faster than FLTrust.

\begin{table}[t]
% \vspace{-0.1in}
\caption{Testing accuracy, FPR, and FNR of different malicious client detection methods under different attacks. The experimental results of targeted attacks are in the form of ``testing accuracy/attack success rate.''}
\label{tab:detect_mal}
\vspace{-0.05in}
% \footnotesize
\centering
    \scalebox{0.66}{
    \begin{tabular}{|c|c|ccc|ccc|ccc|} %第一列设置宽度为45pt 全为左对齐 没有分割线
		%\setlength{\tabcolsep}{20mm}
		%\hline  % 表格的横线
		\hline % 顶部线
		  \multirow{2}{*}{\makecell{Dataset\\(Model)}}&\multirow{2}{*}{Attack}& \multicolumn{3}{c|}{Testing accuracy} & \multicolumn{3}{c|}{FPR}& \multicolumn{3}{c|}{FNR}\\
		  \cline{3-11} % & \clien{4-5}
		  & &  Tolpegin & FLDetector & \textbf{\sys} &  Tolpegin & FLDetector & \textbf{\sys} &  Tolpegin & FLDetector & \textbf{\sys}\\%[3pt]只改一行
		\hline  % 表格的横线
		% \multirow{7}{*}{\makecell{Fashion\\-MNIST\\(CNN)}} &None&	\textbf{ 0.89} & 0.86&	 \textbf{0.89}  & / & / & /  & / & /  & /\\
  %       &LF&	 \textbf{0.89} & 0.86 &	 \textbf{0.89}  &  0.11\% & 0.05\% & \textbf{0.00\%} & 0.38\% &\textbf{0.00\%} & 0.02\%\\
  %       &Min-Max  &	 0.88 &  0.08&	 \textbf{0.89}  &  0.38\% & 96.94\%& \textbf{0.27\%} & 0.80\% & 100\% & \textbf{0.40\%}\\
  %       &Min-Sum  &	 0.64 &0.38&	 \textbf{0.89}  &  41.1\% & 100\%& \textbf{0.26\%} & 84.8\% & 100\% & \textbf{0.00\%}\\
  %       &Fang-Trim  &	 0.88 &0.86&	 \textbf{0.89}  &  0.00\% &0.00\% & 0.00\% & 0.00\% & 0.00\% & 0.00\%\\
  %       &Fang-Krum  &	 0.85 &0.10&	 \textbf{0.89}  &  28.2\% &99.46\%&\textbf{0.20\%} & 60.4\% & 100\% & \textbf{0.00\%}\\ 
  %       & Scaling & 0.89 / 0.10 & 0.89 / 0.10 & 0.89 / 0.10 & 0.00\% & 0.00\% & 0.00\% & 0.03\% & 0.00\% & 0.00\%\\ 
        
  %       \hline \hline
        
		\multirow{8}{*}{\makecell{CIFAR10\\(ResNet20)}} &None&	 0.76 &0.75&	0.76  & 0.00\% & 18.2\% & 0.00\% & / & / & / \\
        &LF&	 0.70 &0.72&	 \textbf{0.76}  & 13.1\% & \textbf{0.03\%} & 4.72\% & 19.9\% & 25.53\% & \textbf{2.60\%} \\
        &Min-Max  &	 0.61 &0.11&	 \textbf{0.77}  & 36.5\% & 100\% &\textbf{ 0.00\%} & 88.0\% & 100\% & \textbf{0.00\%} \\
        &Min-Sum  &	 0.59 &0.31&	 \textbf{0.77}  & 38.8\% & 100\% & \textbf{0.00\%} & 78.0\% & 100\% & \textbf{0.00\%} \\
        &Fang-Trim  &	 0.76 &0.74&	 0.76  & 0.00\% & 0.03\% & 0.00\% & 0.00\% & 0.00\% & 0.00\% \\
        &Fang-Krum  &	 0.17 &0.31&	 \textbf{0.77}  & 37.4\% & 87.18\% & \textbf{0.00\%} & 84.0\% & 100\% & \textbf{0.00\%}\\
		& Scaling & 0.76/0.10 &	0.74/0.10 &	\textbf{0.77/0.11} &	0.00\% &	0.00\% &	0.00\% &	0.10\% &	0.00\% &	0.00\% \\
        & DBA & 0.65/0.47 &	0.77/0.10 &	0.77/0.10 &	0.00\% &	0.00\% &	0.00\% &	46.1\% &	0.00\% &	0.00\% \\
        \hline 
  %       \hline
        
		% \multirow{7}{*}{\makecell{CIFAR-100\\(ResNet20)}} &None&  \\
  %       &LF& \\
  %       &Min-Max  & \\
  %       &Min-Sum  & \\
  %       &Fang-Trim  & \\
  %       &Fang-Krum  & \\ 
  %       & Scaling & \\
        
  %       \hline
	\end{tabular}
}
\vspace{-0.1in}
\end{table}
\subsection{The Significance of Learnable Masks}
\label{sec:mask_sig}
We conduct comparative experiments on \sys and the other malicious client detection algorithms, \ie, Tolpegin defense~\cite{tolpegin2020data} and FLDetector~\cite{zhang2022fldetector}. We sample the result of malicious client detection every ten communication rounds and calculate all the metrics described in \S\ref{sec:metrics}. 

In Table \ref{tab:detect_mal}, we can see that FLDetector only works under LF attack, Fang-Trim attack and targeted attack. Min-Max and Min-Sum attacks make FLDetector completely confuse the malicious model updates with the benign model updates.
Tolpegin defense only survives under Fang-Trim attack, and all other attacks greatly harm it. The large FNR of Tolpegin defense means many malicious clients complete their attacks. The large FPR of Tolpegin defense shows that the remaining benign clients may lose some representation, so Tolpegin defense achieves a lousy performance. Our \sys achieves the same prediction accuracy as a no-attack case, with FPR less than 5\% and FNR less than 3\%.

\begin{figure}
 \vspace{-0.1in} 
  \begin{minipage}[t]{.56\linewidth}
    \centering
    \subfigure[Fang-Krum attack]{\label{fig:detect_a}
    \includegraphics[width=0.47\linewidth]{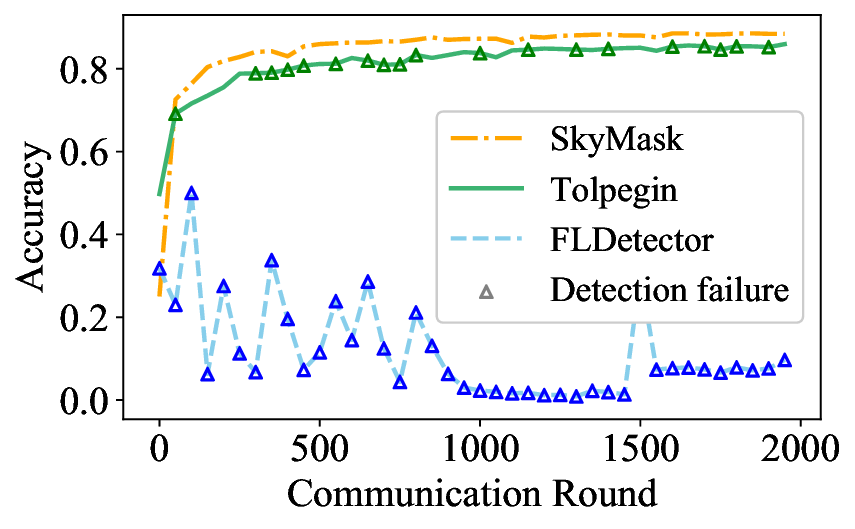}}
    \subfigure[Min-Sum attack]{\label{fig:detect_b}
    \includegraphics[width=0.47\linewidth]{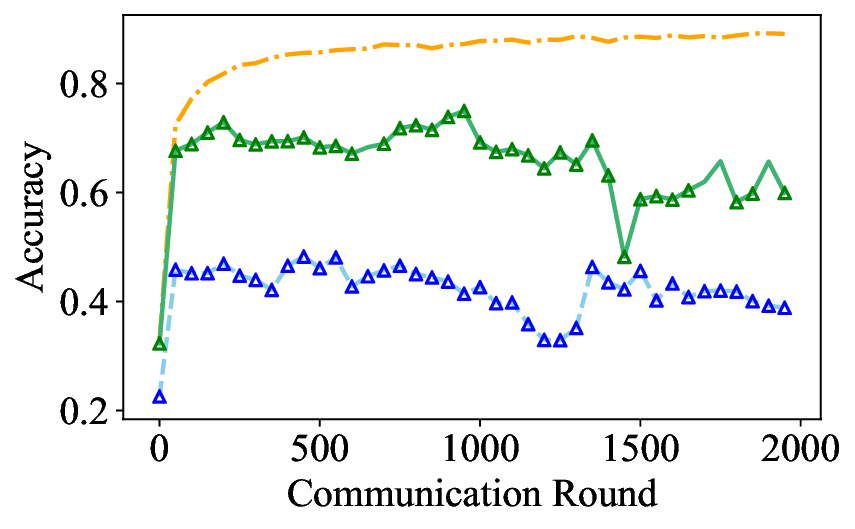}}
    \vspace{-0.1in}
    \caption{The impact of detection failure on different malicious client detection methods.}% and the training efficiency.}% of Tolpegin defense, FLDetector and \textsc{SkyMask}.}
    \label{fig:detect_together}
	% \vspace{-0.2in} 
  \end{minipage} %\hfill
  \begin{minipage}[t]{.42\linewidth}
    \centering
	\vspace{-0.2in}
    \captionof{table}
      {The impact of the total number of clients. (The first line in table is the testing accuracy result of FedAvg under no attack.) % 表格标题
  	\label{tab:scale}	
      }
    \centering\scalebox{0.6}{   
	\begin{tabular}{ccccccc}
		\hline % 
		
        \multirow{2}{*}{\makecell{CIFAR-10\\(ResNet20)}} & \multicolumn{2}{c}{Test Acc.} & \multicolumn{2}{c}{FPR} & \multicolumn{2}{c}{FNR} \\
		\cline{2-7}
		 & 200  &  500  &  200  &  500 &  200  &  500\\%[3pt]只改一行
		\hline  % 表格的横线 \makecell{\textbf{w/o attack} \\ \textbf{(FedAvg)}}
        \textbf{FedAvg} &	0.75 & 0.73 & / & / & / & / \\
        None &	0.75 & 0.73 & / & /& / & /\\
        LF  & 0.75 & 0.72 & 2.94\% & 6.36\% & 3.62\% & 8.36\% \\
        Min-Max & 0.75 & 0.74 & 0.00\% & 0.00\% & 0.00\% & 0.00\%\\
        Min-Sum & 0.74 & 0.73 & 0.00\% & 0.00\% & 0.00\% & 0.00\%\\
        Fang-Trim &	0.75 & 0.73 & 0.00\% & 0.00\% & 0.00\% & 0.00\%\\
        Fang-Krum & 0.75 & 0.72 & 0.00\% & 0.00\% & 0.00\% & 0.00\%\\
        Scaling & 0.75/0.09 & 0.74/0.11 & 0.00\% & 0.00\% & 0.00\% & 0.00\%\\
        DBA & 0.75/0.10 & 0.73 /0.10 & 0.00\% & 0.00\% & 0.00\% & 0.00\%\\
		\hline 
  
	\end{tabular}}
	% \vspace{-0.2in} 
  \end{minipage}
\vspace{-0.4in} 
\end{figure}

% CNN + Fashion-MNIST
% In the Fashion-MNIST part of Table \ref{tab:detect_mal}, we can see that it is hard for Tolpegin defense to detect malicious clients stably, especially under Fang-Krum attack and Min-Sum attack. Its FPR and FNR are pretty high, so it fails to defend against attacks effectively and loses a lot of benign information. Quite the opposite, \sys perfectly completes the detection task and achieves an FNR of less than 4\textperthousand~ on Fashion-MNIST dataset with a CNN model, which means that the server hardly misses any malicious client, and it can eliminate the harm of attacks. \sys also achieves an FPR of less than 3\textperthousand~, so it can preserve most benign clients and better characterize the global data distribution.
% \todo{Legend Here!!!}
Fig.~\ref{fig:detect_a} and Fig.~\ref{fig:detect_b} show the trend of testing accuracy on Fashion-MNIST with a four-layer CNN global model under Fang-Krum and Min-Sum attack. Detection failure means that the FL defense system does not detect all the malicious clients in that round.
% We can see that \sys gradually rises and converges without detection failure. 
Under Fang-Krum attack, Tolpegin defense converges but yields poor testing accuracy, while FLDetector is rendered ineffective. Additionally, both FLDetector and Tolpegin defense exhibit fluctuating and low testing accuracy under Min-Sum attack due to detection failures. In contrast, \sys stably converges without encountering any detection failures.
% Tolpegin defense converges but with poor testing accuracy under Fang-Krum attack and FLDetector is completely damaged. 
% FLDetector and Tolpegin defense exhibit fluctuating and low testing accuracy under Min-Sum attack due to detection failures, while \sys consistently converges without experiencing any detection failures.
% FLDetector and Tolpegin defense are damaged under Min-Sum attack. Each detection failure causes them a fluctuation in testing accuracy, while \sys stably converges without detection failure.

% \vspace{-0.2in} 

\subsection{\textsc{SkyMask}'s Scalability}
To demonstrate the scalability of \sys, we conduct experiments utilizing a ResNet20 global model trained on CIFAR-10. We assess the prediction accuracy of the main task, along with the FPR and FNR of \sys under various attack scenarios, encompassing 200 and 500 clients. %The results of 100 clients are in Table~\ref{tab:detect_mal}. %We have mentioned in \S\ref{sec:dis-dataset} that the local datasets assigned to the clients were disjoint and each client's dataset should be reasonably sized, so for the CIFAR-10 experiments with fewer data samples, we select 200 and 500 clients.

% Table \ref{tab:scale} shows that \sys can work well even if the number of clients increases to 1000 on Fashion-MNIST. As the number of clients increases, more malicious clients are missed under Min-Sum and Fang-Krum attack, but the FNR is still lower than 1.6\%. In addition, the missed malicious local models do not affect the performance of the global model. We can see that the global model of \sys achieves the same testing accuracy as the unattacked global model's obtained by FedAvg.

In Table \ref{tab:scale}, we observe a decrease in the testing accuracy results of the unattacked FedAvg as the total number of clients increased. This can be attributed to the challenges posed by the smaller size of the local datasets and the larger number of clients in the FL training process~\cite{kamp2023federated}. For the LF attack, \sys achieves comparable testing accuracy to the unattacked FedAvg, despite a slight increase in the FNR to approximately 9\%. For all fine-grained attacks and targeted attacks, \sys maintains FPR and FNR at 0\%. 
% Our \sys algorithm demonstrates robust performance under all attacks.
% Therefore, \sys can work well when the total number of clients increases. 
Our \sys algorithm demonstrates robust performance under all attacks even when the total number of clients significantly increases.
%Even if there are millions of clients in the FL system, the server can randomly choose a fixed number (\eg, 100 or 500) of clients in each communication round to execute malicious client detection and aggregation. % The details we will not describe in this paper.

% \input{Table/3-scalability}
\section{Conclusion}
We propose a new attack-agnostic robust FL framework called \sys to defend against Byzantine attacks.
By training parameter-level learnable binary masks on a clean root dataset, \sys is the first to design a fine-grained detection of the poisoned elements of local model updates. 
%
% At the same time, \sys keeps a low FPR and does not affect the normal training process.
% Because \sys analyzes the mask as a whole, it avoids accidentally removing outliers caused by data heterogeneity on some elements of model updates and precisely removes the malicious clients. 
%
Extensive experiments on non-IID datasets under different attacks prove the effectiveness, generality, and scalability of \sys. Our \sys shows a solid defensive capability, better than various robust aggregation algorithms and existing malicious client detection methods. Moreover, our \sys also tackles the problem that most clients are malicious and can defend against attacks with a fraction of malicious clients up to $80\%$. 

% When the system can ensure that the number of benign clients is larger than that of malicious, our \textsc{SkyMask}-NR scheme can also work well with lower overhead. Our method achieves the same or very similar performance as the unattacked FedAvg. 

% ---- Acknowledgment ----
\section*{Acknowledgements}
This work was partially supported by the National Key R\&D Program of China (2022YFB4402102), the Shanghai Key Laboratory of Scalable Computing and Systems, the HighTech Support Program from STCSM (No.22511106200), and Intel Corporation (UFunding 12679). Tao Song is the corresponding author.

\bibliographystyle{splncs04}
\bibliography{reference.bib}

\begin{thebibliography}{10}
\providecommand{\url}[1]{\texttt{#1}}
\providecommand{\urlprefix}{URL }
\providecommand{\doi}[1]{https://doi.org/#1}

\bibitem{bagdasaryan2020backdoor}
Bagdasaryan, E., Veit, A., Hua, Y., Estrin, D., Shmatikov, V.: {How to Backdoor Federated Learning}. In: Proc. AISTATS (2020)

\bibitem{baruch2019little}
Baruch, G., Baruch, M., Goldberg, Y.: {A Little is Enough: Circumventing Defenses for Distributed Learning}. In: Proc. NeurIPS (2019)

\bibitem{bhagoji2019analyzing}
Bhagoji, A.N., Chakraborty, S., Mittal, P., Calo, S.: {Analyzing Federated Learning Through an Adversarial Lens}. In: Proc. ICML (2019)

\bibitem{blanchard2017machine}
Blanchard, P., El~Mhamdi, E.M., Guerraoui, R., Stainer, J.: {Machine Learning with Adversaries: Byzantine Tolerant Gradient Descent}. In: Proc. NeurIPS (2017)

\bibitem{cao2020fltrust}
Cao, X., Fang, M., Liu, J., Gong, N.Z.: {FLTrust: Byzantine-Robust Federated Learning via Trust Bootstrapping}. In: Proc. NDSS (2021)

\bibitem{dean2012distribute}
Dean, J., Corrado, G., Monga, R., Chen, K., Devin, M., Mao, M., Ranzato, M.a., Senior, A., Tucker, P., Yang, K., Le, Q., Ng, A.: {Large Scale Distributed Deep Networks}. In: Proc. NeurIPS (2012)

\bibitem{fang2020local}
Fang, M., Cao, X., Jia, J., Gong, N.: {Local Model Poisoning Attacks to Byzantine-Robust Federated Learning}. In: Proc. USENIX Security (2020)

\bibitem{guerraoui2018hidden}
Guerraoui, R., Rouault, S., et~al.: {The Hidden Vulnerability of Distributed Learning in Byzantium}. In: Proc. ICML (2018)

\bibitem{guo2021siren}
Guo, H., Wang, H., Song, T., Hua, Y., Lv, Z., Jin, X., Xue, Z., Ma, R., Guan, H.: Siren: Byzantine-robust federated learning via proactive alarming. In: Proc. SoCC (2021)

\bibitem{guo2021multi}
Guo, P., Wang, P., Zhou, J., Jiang, S., Patel, V.M.: Multi-institutional collaborations for improving deep learning-based magnetic resonance image reconstruction using federated learning. In: Proc. CVPR (2021)

\bibitem{hsu2020federated}
Hsu, T.M.H., Qi, H., Brown, M.: Federated visual classification with real-world data distribution. In: Proc. ECCV (2020)

\bibitem{kamp2023federated}
Kamp, M., Fischer, J., Vreeken, J.: Federated learning from small datasets. In: Proc. ICLR (2023)

\bibitem{krizhevsky2009learning}
Krizhevsky, A., Hinton, G., et~al.: {Learning Multiple Layers of Features from Tiny Images}. Technical Report  (2009)

\bibitem{li2021fedmask}
Li, A., Sun, J., Zeng, X., Zhang, M., Li, H., Chen, Y.: Fedmask: Joint computation and communication-efficient personalized federated learning via heterogeneous masking. In: Proc. SenSys (2021)

\bibitem{li20233dfed}
Li, H., Ye, Q., Hu, H., Li, J., Wang, L., Fang, C., Shi, J.: 3dfed: Adaptive and extensible framework for covert backdoor attack in federated learning. In: Proc. IEEE S\&P (2023)

\bibitem{li2023martfl}
Li, Q., Liu, Z., Li, Q., Xu, K.: martfl: Enabling utility-driven data marketplace with a robust and verifiable federated learning architecture. In: Proc. CCS (2023)

\bibitem{li2023diverse}
Li, Y., Wang, X., Yang, L., Feng, L., Zhang, W., Gao, Y.: Diverse cotraining makes strong semi-supervised segmentor. In: ICCV (2023)

\bibitem{liu2023clip}
Liu, J., Zhang, Y., Chen, J.N., Xiao, J., Lu, Y., A~Landman, B., Yuan, Y., Yuille, A., Tang, Y., Zhou, Z.: Clip-driven universal model for organ segmentation and tumor detection. In: Proc. CVPR (2023)

\bibitem{liu2021feddg}
Liu, Q., Chen, C., Qin, J., Dou, Q., Heng, P.A.: Feddg: Federated domain generalization on medical image segmentation via episodic learning in continuous frequency space. In: Proc. CVPR (2021)

\bibitem{liu2020fedvision}
Liu, Y., Huang, A., Luo, Y., Huang, H., Liu, Y., Chen, Y., Feng, L., Chen, T., Yu, H., Yang, Q.: Fedvision: An online visual object detection platform powered by federated learning. In: Proc. AAAI (2020)

\bibitem{lyu2023poisoning}
Lyu, X., Han, Y., Wang, W., Liu, J., Wang, B., Liu, J., Zhang, X.: Poisoning with cerberus: stealthy and colluded backdoor attack against federated learning. In: Proc. AAAI (2023)

\bibitem{mcmahan2017communication}
McMahan, B., Moore, E., Ramage, D., Hampson, S., y~Arcas, B.A.: {Communication-Efficient Learning of Deep Networks from Decentralized Data}. In: Proc. AISTATS (2017)

\bibitem{miao2022privacy}
Miao, Y., Liu, Z., Li, H., Choo, K.K.R., Deng, R.H.: Privacy-preserving byzantine-robust federated learning via blockchain systems. IEEE Transactions on Information Forensics and Security  \textbf{17},  2848--2861 (2022)

\bibitem{nguyen2022flame}
Nguyen, T.D., Rieger, P., De~Viti, R., Chen, H., Brandenburg, B.B., Yalame, H., M{\"o}llering, H., Fereidooni, H., Marchal, S., Miettinen, M., et~al.: $\{$FLAME$\}$: Taming backdoors in federated learning. In: Proc. USENIX Security (2022)

\bibitem{nguyen2024iba}
Nguyen, T.D., Nguyen, T.A., Tran, A., Doan, K.D., Wong, K.S.: Iba: Towards irreversible backdoor attacks in federated learning. Proc. NeurIPS  (2024)

\bibitem{qureshi2021performance}
Qureshi, N.B.S., Kim, D.H., Lee, J., Lee, E.K.: On the performance impact of poisoning attacks on load forecasting in federated learning. In: Proc. UbiComp (2021)

\bibitem{rieger2022deepsight}
Rieger, P., Nguyen, T.D., Miettinen, M., Sadeghi, A.R.: Deepsight: Mitigating backdoor attacks in federated learning through deep model inspection. In: Proc. NDSS (2022)

\bibitem{shejwalkar2021manipulating}
Shejwalkar, V., Houmansadr, A.: {Manipulating the Byzantine: Optimizing Model Poisoning Attacks and Defenses for Federated Learning}. In: Proc. NDSS (2021)

\bibitem{sun2019can}
Sun, Z., Kairouz, P., Suresh, A.T., McMahan, H.B.: {Can You Really Backdoor Federated Learning?} arXiv preprint arXiv:1911.07963  (2019)

\bibitem{tolpegin2020data}
Tolpegin, V., Truex, S., Gursoy, M.E., Liu, L.: {Data Poisoning Attacks against Federated Learning Systems}. In: Proc. ESORICS (2020)

\bibitem{wang2022protect}
Wang, J., Guo, S., Xie, X., Qi, H.: Protect privacy from gradient leakage attack in federated learning. In: INFOCOM (2022)

\bibitem{xie2019dba}
Xie, C., Huang, K., Chen, P.Y., Li, B.: {DBA: Distributed Backdoor Attacks against Federated Learning}. In: Proc. ICLR (2019)

\bibitem{yin2018byzantine}
Yin, D., Chen, Y., Kannan, R., Bartlett, P.: {Byzantine-Robust Distributed Learning: Towards Optimal Statistical Rates}. In: Proc. ICML (2018)

\bibitem{zhang2024a3fl}
Zhang, H., Jia, J., Chen, J., Lin, L., Wu, D.: A3fl: Adversarially adaptive backdoor attacks to federated learning. In: Proc. AAAI (2024)

\bibitem{zhang2021federated}
Zhang, L., Luo, Y., Bai, Y., Du, B., Duan, L.Y.: Federated learning for non-iid data via unified feature learning and optimization objective alignment. In: ICCV (2021)

\bibitem{zhang2022fldetector}
Zhang, Z., Cao, X., Jia, J., Gong, N.Z.: Fldetector: Defending federated learning against model poisoning attacks via detecting malicious clients. In: Proc. KDD (2022)

\bibitem{zhou2019deconstructing}
Zhou, H., Lan, J., Liu, R., Yosinski, J.: Deconstructing lottery tickets: Zeros, signs, and the supermask. In: Proc. NeurIPS (2019)

\end{thebibliography}

\end{document}